\documentclass[conference]{IEEEtran}
\IEEEoverridecommandlockouts
\usepackage{cite}
\usepackage{amsmath,amssymb,amsfonts}
\usepackage{algorithmic}
\usepackage{graphicx}
\usepackage{textcomp}
\usepackage{xcolor}
\def\BibTeX{{\rm B\kern-.05em{\sc i\kern-.025em b}\kern-.08em
    T\kern-.1667em\lower.7ex\hbox{E}\kern-.125emX}}
\begin{document}


\title{On eliminating blocking interference of RFID unauthorized reader detection system\\
}

\author{\IEEEauthorblockN{1\textsuperscript{st} Degang Sun}
\IEEEauthorblockA{\textit{School of Cyber Security (of UCAS)} \\
Beijing, China \\
sundegang@iie.ac.cn}
\and
\IEEEauthorblockN{2\textsuperscript{nd} Yue Cui}
\IEEEauthorblockA{\textit{Institute of Information Engineering (of CAS)} \\
\textit{School of Cyber Security (of UCAS)}\\
Beijing, China \\
cuiyue@iie.ac.cn}
\and
\IEEEauthorblockN{3\textsuperscript{rd} Siye Wang}
\IEEEauthorblockA{\textit{Institute of Information Engineering (of CAS)} \\
\textit{School of Cyber Security (of UCAS)}\\
Beijing, China \\
wangsiye@iie.ac.cn}
\and
\IEEEauthorblockN{4\textsuperscript{th} Yanfang Zhang}
\IEEEauthorblockA{\textit{Institute of Information Engineering (of CAS)} \\
\textit{School of Cyber Security (of UCAS)}\\
Beijing, China \\
zhangyanfang.ac.cn}
}

\maketitle

\begin{abstract}
RFID as an important component technology of IoT faces important security risks while being rapidly applied, among which the discovery of unauthorized readers in space is crucial. There are some researches proposed the unauthorized reader detection algorithm based on commercial off the shell(COTS) devices, but these detection algorithms are often easily affected by moving objects blocking interference in space, causing false alarms. We propose a new method of eliminating moving object interference, which can reduce the system false alarm rate to less than 7.9\% by experimental testing.
\end{abstract}


\section{Introduction}
Due to the reader is the main way to obtain the UHF RFID air interface data, and in the ISO 18000-6C protocol\cite{b1} for RFID system used by the COTS devices lacks of security certification between the tags and the reader, so that any unauthorized reader in the space can communicate with the tags and get access to the tags data, resulting in data leakage, data tampering and other security threats\cite{b2}.

And for the security of commercial RFID system, there are many researchers' related work, such as: Pramod present physical unclonable function based unilateral authentication protocol for RFID system\cite{b3}. The algorithm has been implemented on an 8-bit open-loop resonator based chipless RFID tag based system and is validated using BASYS 2 FPGA board based platform \cite{b4}. Security analysis shows that SLAP guarantees the functionalities of mutual authentication as well as resistance to various attacks such as de-synchronization attack, replay attack and traceability attack, etc \cite{b5}. But taking the cost, scalability into consideration, using the commercial reader itself as monitoring equipment, and does not impose too much encryption calculation, through its signal when normal communication with the tags  to discern whether there is an unauthorized reader is the best way. Commercial reader can obtain the data like, RSSI, Phase, time stamp, throughput rate. Using these indicators such as \cite{b6} and \cite{b7}.  When using these methods in the office environment, we found that the surrounding pedestrian walking or blocking will affect the signal, causing false alarm, so this paper introduces how to eliminate the blocking interference of these moving objects.
\section{unauthorized reader detection system analysis}
\begin{figure}[h] 
    \centering
    \includegraphics[scale=0.4]{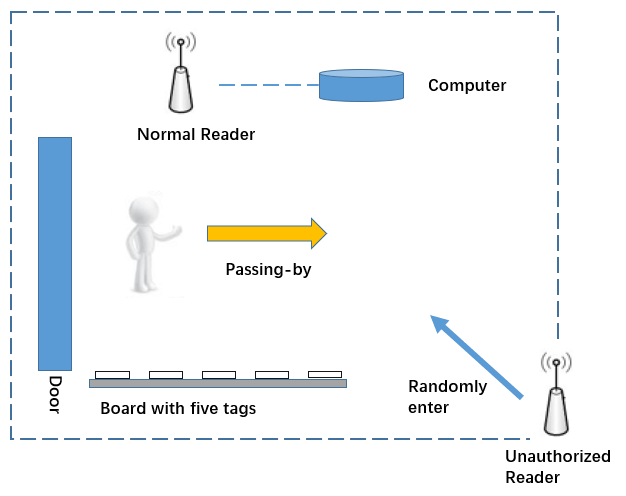}
    \caption{Experiment setup}
    \label{fig:experiment}
\end{figure}

In this section, we will analyze the RFID unauthorized reader(UR) detection system based on ASTI(adjacent signals time interval) which is a optimal parameter comparing to throughput rate because lower time delay\cite{b6}. In addition, according to the needs of actual use, we chose a common scenario to arrange the experiment - placing the reader at the entrance(as \ref{fig:experiment}).And when human or other object passes by, how the performance of UR detection system changes.

\subsection{UR detection system}
\begin{figure}[h] 
    \centering
    \includegraphics[scale=0.55]{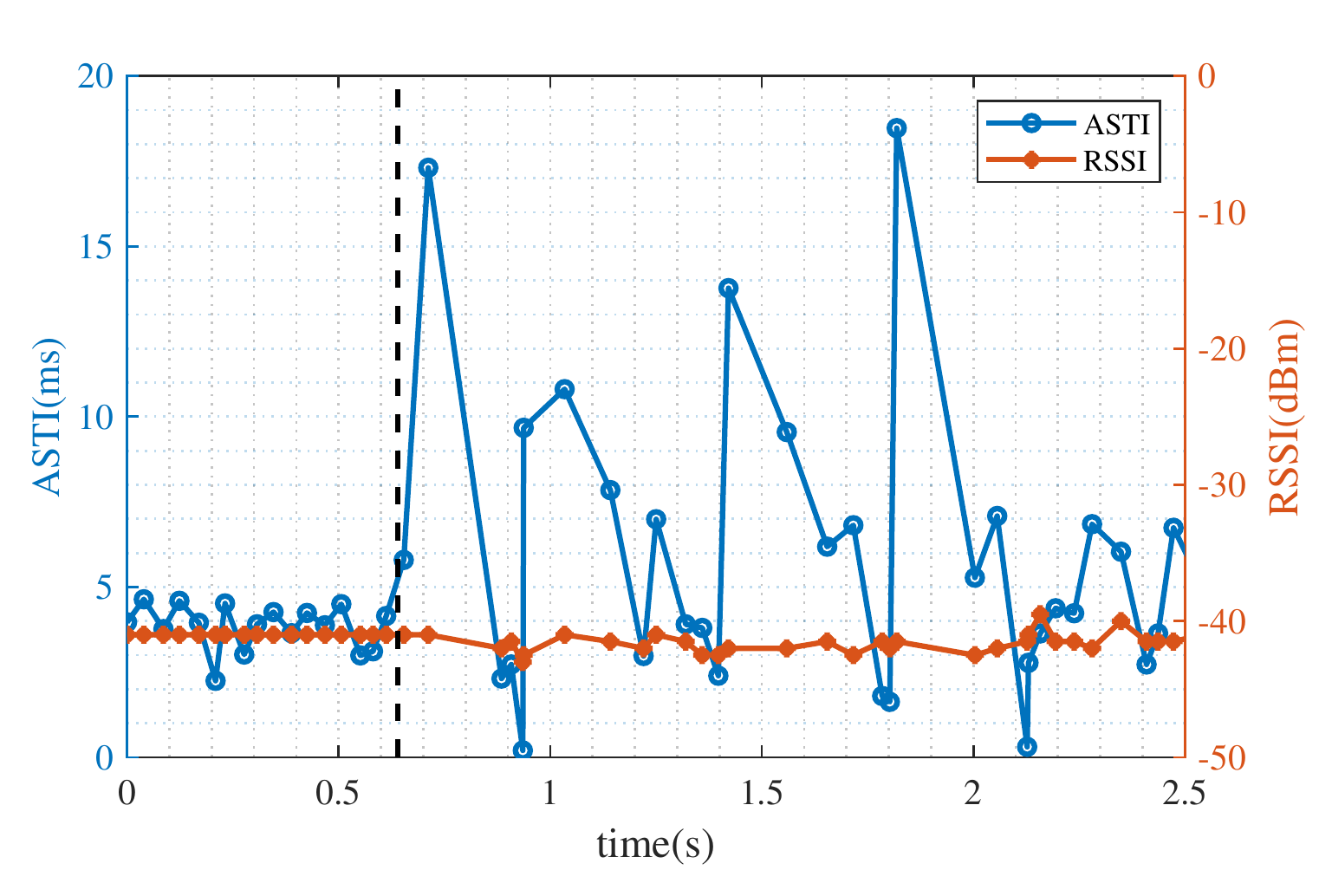}
    \caption{ASTI \& RSSI under unauthorized reader}
    \label{fig:urasti}
\end{figure}

We setup the pre-experiment as \ref{fig:experiment},the normal reader is set at the entrance and start up to communicate with the five tags equally spaced at the board in front of the reader. Then the UR randomly enters and start. Figure\ref{fig:urasti} shows one of the results,the black line indicates the time UR starts. ASTI is calculated by:
\begin{equation}
   ASTI=Timestamp(i)-Timestamp(i-1)
\end{equation}

It's obvious that after UR starting, the variation of ASTI is significant but the RSSI varies as small fluctuations. So the change rate of ASTI can be adopted as a parameter for UR detection. We adopt the detection algorithm in\cite{b7}.

\begin{figure}[h] 
    \centering
    \includegraphics[scale=0.55]{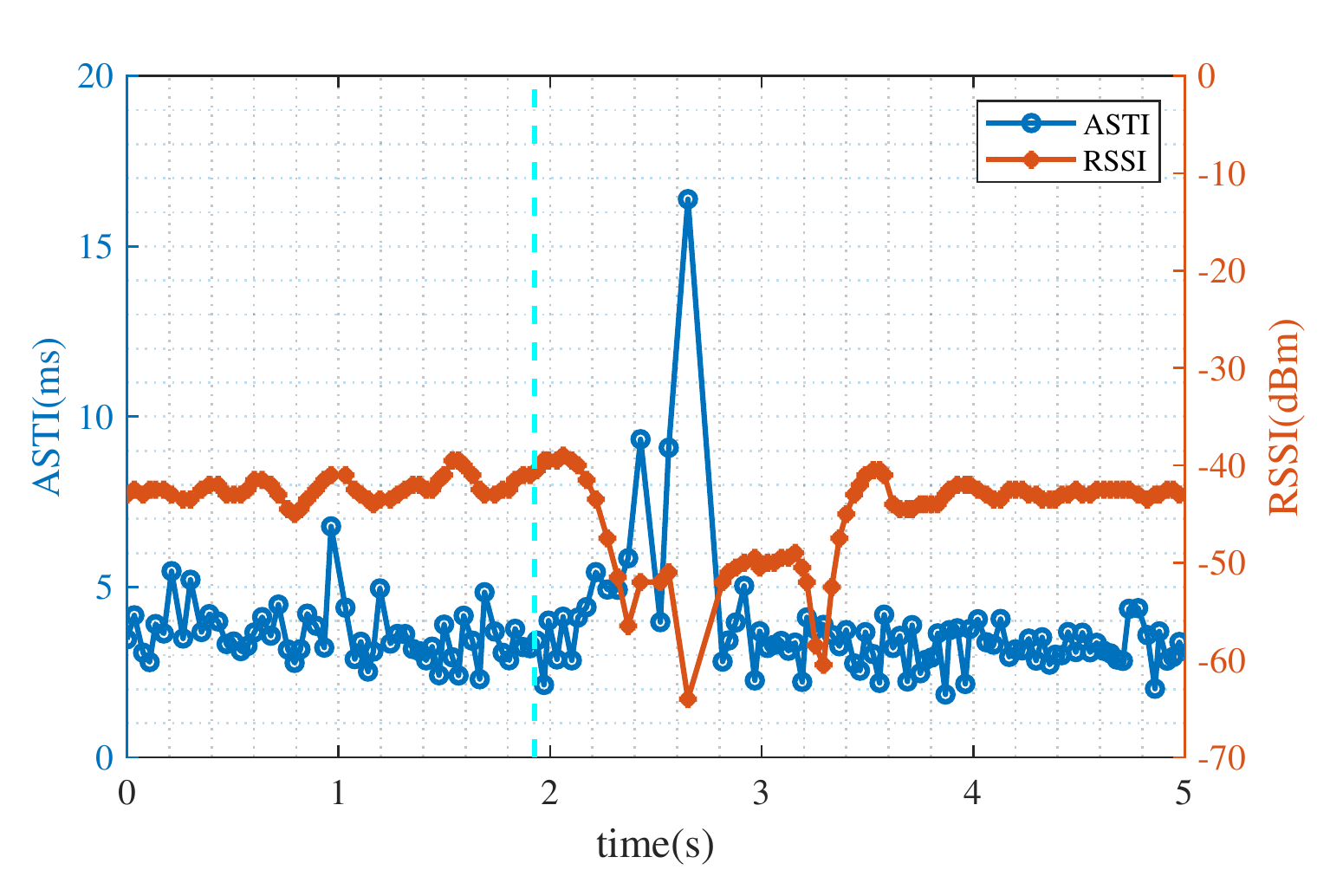}
    \caption{ASTI \& RSSI under unauthorized reader and passing-by}
    \label{fig:objasti}
\end{figure}

In the next part, we start the normal reader and UR, then a man walk through the entrance to simulator the true environment for daily use. Figure \ref{fig:objasti} shows one of the experiment results. The light blue line indicates when the man start to block the tag and normal reader.

From the result, we can find that ASTI also changes noticeable as human passing so does the RSSI. But because the RSSI is not sensitive to whether UR is at present, we can use RSSI as the parameter for human-centered computing.

Human passing-by and UR cause the same result for the ASTI UR detection so the false-alarm rate(FPR) will be high in the real environment. It's vital to find the way to eliminate the interference of human.

\subsection{Model of moving human}
First we model and analyze the action of a moving object passing a doorway using pedestrian data as an example. Suppose someone with top view length and width $l_h$ and $d_h$ respectively, passes the doorway from the middle with speed $V$m/s, the tag is equally spaced $d_{set}$ placed, the effective length of the signal reception of the tag is $l_{tagi}$, the red line part is the angle of the human body completely blocking the sight distance path, the corresponding impact distance of $d_{in}$ and impact time $T_{in}$ can be found as follows:
\begin{equation}
    d_{in}=\frac{\tfrac{1}{2}(d_{door}+l_h)}{tan(\frac{\pi}{2}-\theta_{tag1B})}-
    \frac{\tfrac{1}{2}(d_{door}-l_h)}{tan(\frac{\pi}{2}-\theta_{tag1A})}+d_h
    \end{equation}

\begin{equation}
    T_{in}=\frac{d_{in}}{V}
\end{equation}

\begin{figure}[h] 
    \centering
    \includegraphics[scale=0.3]{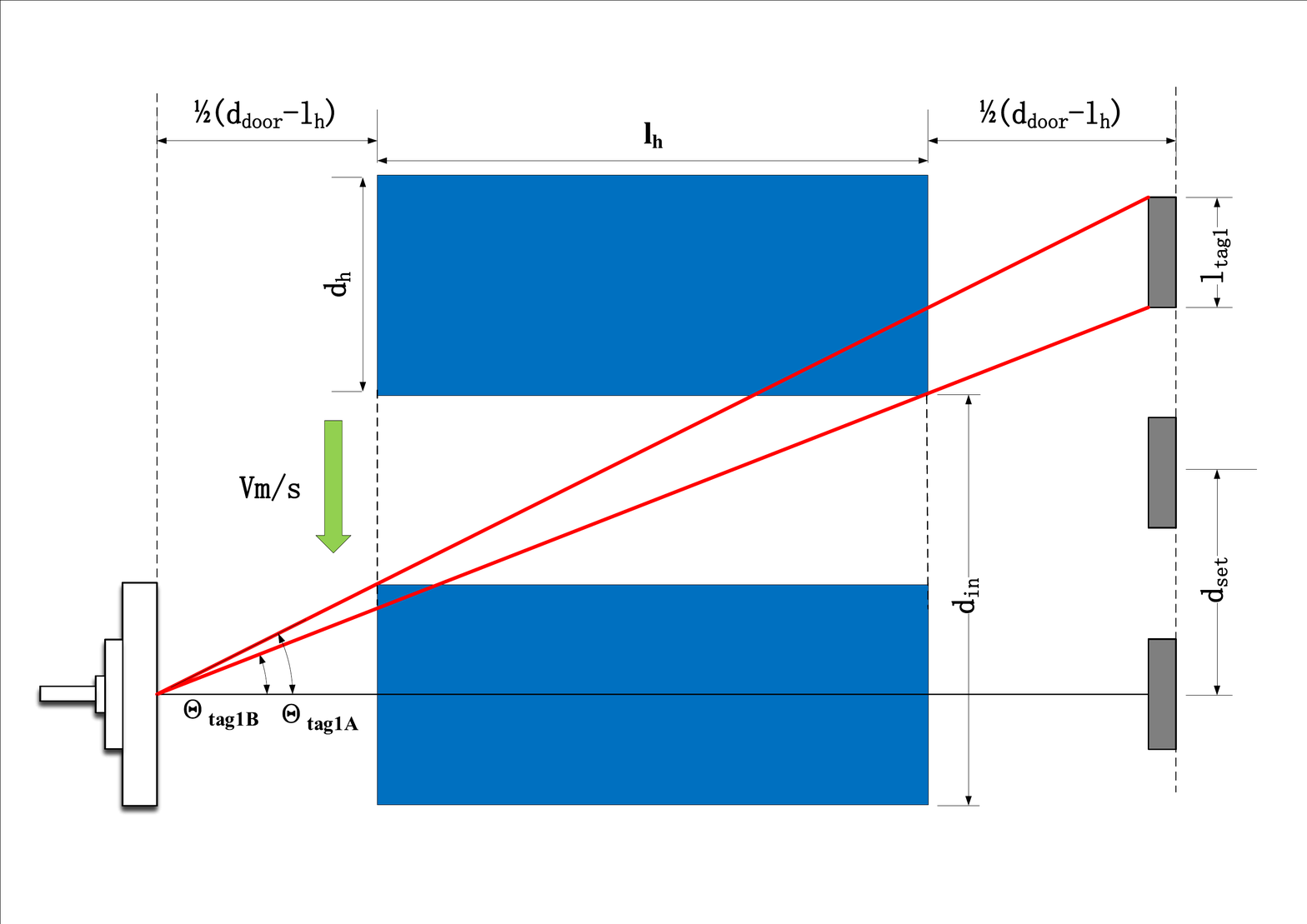}
    \caption{Moving object passing doorway model}
    \label{fig:humanobj}
\end{figure}

In RFID system, the signal received by the reader antenna can be divided into line-of-sight (LOS) signal and multipath effect signal superposition caused by other reflection paths.
\begin{equation}\label{H}
    H(f)=\sum_{n=0}^{N-1} \rho_{n} e^{j \theta_{n}} e^{-j 2 \pi f \tau_{n}}
    \end{equation}
Equation (\ref{H}) represents the superposition of the arrival signals of N paths. where: $ \rho_{n}$ denotes the intensity of the signal; $\theta_{n}$ denotes the phase bit; $\tau_{n}$ denotes the arrival delay. In general, the signal energy of the line-of-sight path (the straight-line path between the sender and the receiver, denoted as n=0 part) ($\rho_{n}$ in Eq.) is much larger than the signal energy of the other paths, and thus dominates in the superposition.
dominant position. In an ideal environment, when multipath effects do not exist, it can be considered that $T_{in} \doteq \Delta t_{tagi}$.
In practice, the tag acquires multivariate data due to environmental multipath and the passage of objects generating new multipath effects
\section{Eliminating blocking interference method}
From the previous experiments and modeling analysis, we figure out the key of getting the time by passing-by is the velocity estimation of the object. So in this section, we will introduce the method for velocity estimation and furthermore eliminating the blocking interference.

\subsection{Data pre-processing}\label{pre}
We use a Moving Average Filter to smooth the acquired data for further processing. Due to its polling mechanism, the RFID reader can only communicate with one tag at a time and the sampling frequency of commercial readers may only reach 30Hz in the environment, however, since the object is always in motion, we want to capture more readings at the same time. Therefore, we use the Hermite Interpolating Polynomial (PCHIP) method to interpolate the data samples at the desired point in time, which allows for more accurate interpolation at a considerably higher efficiency.

\subsection{Velocity estimation based on Pearson correlation coefficient(PCCs)}
At the beginning, a database of moving object speed should be building as the reference for matching. To achieve this, we arranged a dummy placing at a vehicle fixed on the slide rail, which would moving at the set direction with a set velocity. We used the machine to simulate real human or other object passing through the door like \ref{fig:experiment}. The velocity is raised in steps of 0.02 from 0.1m/s to 2m/s and at each velocity tested for 200 times then taking the average number as the result.

When the back end computer newly receives RSSI from the reader, Pearson correlation coefficient\cite{b8} will be used for matching its velocity to the database. Then select the velocity which PCCs is the closest to 1 as the estimation of the velocity of the moving human.

The Pearson correlation coefficient between two variables is defined as the product of the covariance of the two variables divided by their standard deviations:
\begin{equation}\label{P}
    {\displaystyle \rho _{X,Y}={\mathrm {cov} (X,Y) \over \sigma _{X}\sigma _{Y}}={E[(X-\mu _{X})(Y-\mu _{Y})] \over \sigma _{X}\sigma _{Y}}}
    \end{equation}
The closer the correlation coefficient of two variables is to 1, the more significant the linear relationship between the two variables, i.e., the more relevant. One important mathematical property of Pearson's correlation coefficient is that because a change in the position and scale of the two variables does not cause a change in this coefficient, i.e. it the invariance of the change (determined by the sign). That is, if we move $X$ to $a + bX$ and move $Y$ to $c + dY$, where $a, b, c$ and $d$ are constants, it does not change the correlation coefficient of the two variables.

\subsection{Eliminating blocking interference}\label{eli}
After moving speed database building for the velocity estimation goal, when we receive data from the reader, we have to decide whether there is someone passing through causing alarm or real UR existing. 

From the pre-experiment, it can be calculated that the change rate of RSSI with UR is less than 5\%. So it can be used as a threshold to judge object passing-by.

The moment UR detection system send alarm, RSSI change rate is calculated and if the rate is larger than 5\%, velocity estimation part work to match the object speed. Then  $T_{in}$ the time of UR detection influenced add the set window time will be thought as the new window time back to UR detection system:$window=window+T_{in}$. If the system still alarms, We think UR is shown up.
\section{Experiment}
In this section, we arranged some experiment to evaluate our method. The experiment environment was a normal office, devices setup like Figure \ref{fig:experiment}, tested 500 times each group. We choose two Impinj R420 as readers and tags are of H47.
\subsection{Accuracy of UR detection system}
We tested the accuracy of UR detection system in the three conditions: 
\begin{itemize}
    \item without human passing-by:free
    \item with human passing-by:interference
    \item with human passing-by and eliminating method:eliminated
\end{itemize}
The result shows that our method can obviously reduce the interference of human passing-by.
\begin{figure}[h] 
    \centering
    \includegraphics[scale=0.7]{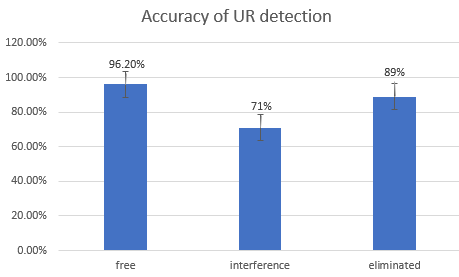}
    \caption{Accuracy of UR detection system}
    \label{fig:accur}
\end{figure}

\subsection{Evaluation of FPR}
\begin{figure}[h] 
    \centering
    \includegraphics[scale=0.6]{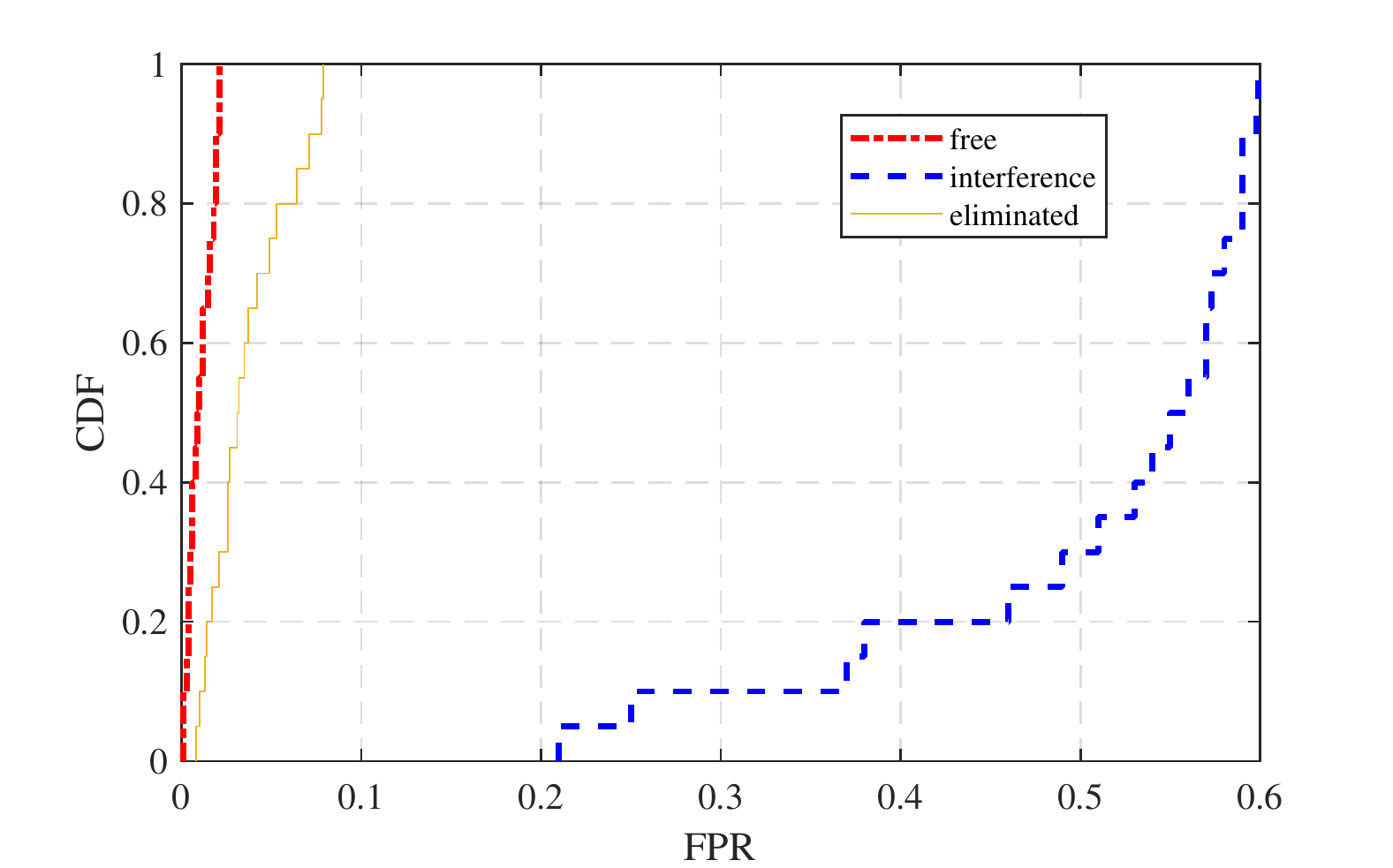}
    \caption{Evaluation of FPR in three conditions}
    \label{fig:CDF}
\end{figure}

In this experiment, we tested for evaluation of FPR which is the most concern of our method. It's can be seen from the Figure\ref{fig:CDF} , our method significantly lower the FPR comparing to the unused one.

\subsection{Accuracy of velocity estimation}
In this experiment, we arranged an experiment for testing velocity estimation part. High accuracy has occupied in all three groups of experiment.

\begin{figure}[h] 
    \centering
    \includegraphics[scale=0.7]{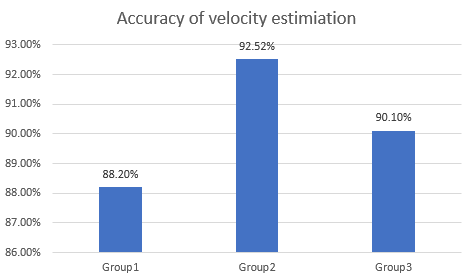}
    \caption{Accuracy of velocity estimation}
    \label{fig:accv}
\end{figure}

\section{conclusion}
This paper proposes a new method of eliminating the human or other object moving in the area causing false alarm for the unauthorized reader detection system. At first, analyze the UR detection system and modeling moving human, then finding key of velocity estimation. Next, by building speed database and using PCCs, put forward the method to eliminate the interference. At last, test our method through experiment and the results show the excellent performance.

\section*{Acknowledgment}
This subject comes from

\vspace{12pt}

\end{document}